\newcommand{\cm}{{~\rm cm}}
\newcommand{\s}{{~\rm s}}

\newcommand{\K}{{~\rm K}}
\newcommand{\erg}{{~\rm erg}}

\newcommand{\keV}{{~\rm keV}}

\documentclass[12pt,preprint]{aastex}
\shortauthors{Sternberg et al.}

\begin{document}

\title{WAS AN OUTBURST OF AQUILA X-1 A MAGNETIC FLARE?}

\author{Noam Soker\altaffilmark{1}}

\altaffiltext{1}{Department of Physics, Technion$-$Israel Institute of
Technology,
Haifa 32000, Israel; soker@physics.technion.ac.il}

\begin{abstract}
I point to an interesting similarity in the radio and the soft X-ray light curves
between the November 2009 outburst of the X-ray binary Aquila~X-1 and some solar flares.
The ratio of the soft X-ray and radio luminosities of Aquila~X-1 in that outburst is also
similar to some weak solar flares, and so is the radio spectrum near 8~GHz.
Based on these, as well as on some other recent studies
that point to some similar properties of accretion disk coronae and stellar flares,
such as ratio of radio to X-ray luminosities (Laor \& Behar 2008),
I speculate that the soft X-ray outburst of Aquila~X-1 was related to a huge
magnetic flare from its disk corona.
\end{abstract}


\section{INTRODUCTION}
\label{sec:intro}

In a recent paper Miller-Jones et al. (2010, here after M2010) present a detail study of
an outburst of the X-ray binary (XRB) Aquila~X-1 (Aql~X-1).
Aql~X-1 is a binary system of a K7V star and a neutron star (NS) that
experiences repeated outbursts.
M2010 discuss the outburst properties in the context of the transition from the hard
state to the soft state and back, and plot the evolution of Aql~X-1 on the
hardness-intensity diagram (HID) commonly used for outbursts of accreting black hole (BH).

{{{{{
During the canonical hard state of a BH XRB outburst, a steady optically
thick compact jet is expected.
During the hard to soft state transition (and sometimes back the other way)
a major radio flare, brighter than the compact jet, occurs, and optically thin
jet ejecta are often observed or inferred to be launched when the ``jet line''
is crossed (e.g., Fender et al. 2009).
It is still not known whether the compact jet exist during the soft state.
M2010 notice two significant differences between the outburst of Aql~X-1 and outbursts of BHs.
These are the flat radio spectrum between 5 and 8~GHz, and the absence of bright,
optically thin, relativistically moving knots.
A major radio outburst is observed to occur during the hard-soft transition in the
outburst of Aql~X-1. However, during this transition, its spectrum
is flat and no jet ejecta are seen.
G2010 argue that during this transition only the compact jet is being observed.
The presence of a compact jet in this transition may or may not be consistent
with BH XRBs.
It is the lack of transient, optically thin jet emission in this transition that is different
from other BH XRB outbursts. }}}}}

In light of these differences and some similarity with solar flares,
I speculate on an alternative interpretation to the radio and soft X-ray emissions
(but not to the hard X-ray emission preceding these two) based on magnetic flares.
The alternative interpretation of magnetic activity was discussed for an accreting
white dwarf by Soker \& Vrtilek (2009).
Soker \& Vrtilek (2009) suggested that the radio emission from an outburst of the
dwarf nova SS Cyg reported by K\"ording et al. (2008) originated from
magnetic activity that formed a corona similar to coronae found in magnetically active
stars, rather than from jets.
Soker \& Vrtilek (2009) based their claim on the results of Laor \& Behar (2008; hereafter LB2008),
who found that when the ratio between radio and X-ray fluxes of accretion disks in
radio-quiet quasars is as in active stars, $L_r/L_x \la 10^{-5}$, then most of the radio
emission might come from coronae.
{{{{{ In LB2008, the radio luminosity is $\nu L_\nu$, usually at around 6 cm,
while $L_x$ is the integrated X-ray luminosity in the range $0.2-20 \keV$. }}}}}
Jets might still occur. If the magnetic activity in erupting accreting disks is
similar to that in active stars, then mass ejection, is expected.
The presence of coronae above accretion disks
(e.g., Galeev et al. 1979; Done \& Osborne 1997; Wheatley \& Mauche 2005),
and the connection between coronae and jets
(e.g., Fender et al. 1999; Markoff et al. 2005; Rodriguez \& Prat 2008) has already been proposed.
However, the results of Ishida et al. (2009; also LB2008)
put the presence of coronae in accretion disks on a solid ground, and further
suggest that magnetic activity similar to that in active stars occur in these coronae.

The speculative interpretation in the present letter (Section \ref{sec:flare}) is based on three properties of the
November 2009 outburst of Aql~X-1 (M2010).
I emphasize that I do not propose an alternative explanation to the hard ($>15 \keV$) X-ray peak,
{{{{{ and in any case postpone its detail study for a future paper. }}}}}
In X-ray transients the hard X-ray peak can generally be accounted for by a disk
instability as studied by Dubus et al. (2001).
In the present case, the hard X-ray peak has a triangular shape (see definition in Chen et al. 1997)
for $\sim 12~$d.
The radio peak appears during the decay phase of the hard X-ray emission.
The soft (RXTE ASM; $2-10 \keV$) X-ray emission appears with the hard X-ray emission, but its large and rapid
rise starts only after the radio peak (see below).
The two peaks, one in the hard X-ray followed by one in the soft X-ray, can be seen for another
flare of Aql~X-1 in Yu et al. (2003), but they do not have radio observations.
The two X-ray peaks do not resemble at all the secondary peaks discussed for X-ray novae
by Chen et al. (1997), and must be explained by a different process.
Among XRB systems the double peak structure of the outburst is common, like XTE J1859+226
(Brocksopp et al. 2002) in which the radio peak occurs at the start of the extended soft X-ray peak;
the hard X-ray peak occurs before the peaks of the radio and of the soft X-ray emissions (Brocksopp et al. 2002).
Such a structure is seen in solar flares as well.
In the magnetic flare model the hard X-ray peak is related to the event that rapidly
amplifies the magnetic field. This field later powers the radio and soft X-ray emissions.

The appearance of the hard X-ray and soft X-ray peaks one after the other
is quite similar to that seen in BH XRB systems,
but it is not at all similar to the delay in rise to maximum between the optical
and extreme UV and X-ray emissions in dwarf novae (Mauche et al. 2001; Wheatley et al. 2003).
In Aql~X-1 the two peaks are separated, while in dwarf novae their behavior with time
is more or less similar, with a relatively short delay.
The disk instability can account for the delay in dwarf novae (Schreiber et al. 2003),
but here a different explanation is required.

\section{THE MAGNETIC FLARE INTERPRETATION}
\label{sec:flare}

\subsection{Light curves}
\label{sec:light}

Many solar flares, as well as of similar stars such as UV Ceti (G\"udel et al. 1996),
show the Neupert effect (Neupert 1968).
This effect is a behavior where the integration of the radio flux (and in many cases
the non-thermal hard X-ray emission) is proportional to the X-ray flux at rise.
In some cases, the radio peak comes at the beginning of the X-ray rise.

\begin{figure}
\vskip +3.5 cm
\includegraphics[scale=0.70]{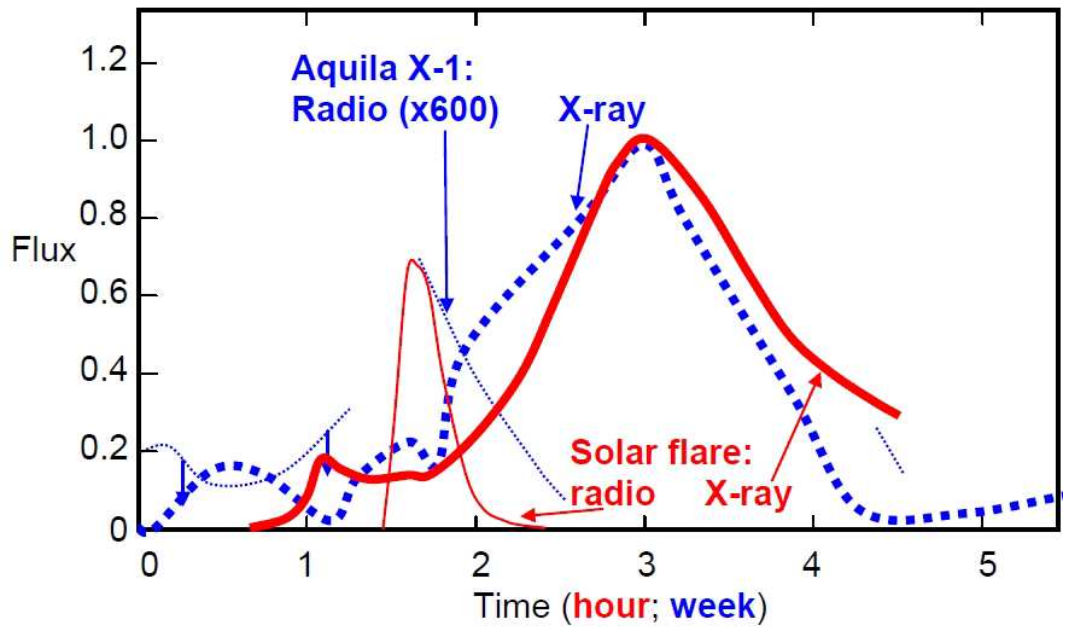}
\caption{The radio (thin lines) and soft X-ray (thick lines; $2-10 \keV$) fluxes in relative units
of Aquila~X-1 November 2009 outburst (from M2010; blue dashed line), and a solar
flare from April 26, 1981 as compiled by G\"udel et al. (1996; red solid line; $0.1-2.4 \keV$).
For Aql~X-1 there are three radio segments, where the left one is only an upper limit.
{{{{{ The ratio of radio to X-ray in Aql~X-1 is $\sim 600$ times smaller than that in the
solar flare. In the plot, the Aql~X-1 radio flux was multiplied by 600. }}}}}
Each time unit is one hour for the solar flare and 7.1 days for Aql~X-1.
}
\label{fig:flare}
\end{figure}

In Figure \ref{fig:flare}, I compare the behavior of one specific solar flare as compiled by
G\"udel et al. (1996; more detail in Cliver et al. 1986, Dennis \& Zarro 1993, and
Benz \& G\"udel 1994),  with the November 2009 outburst of Aql~X-1 (M2010) in radio and soft X-ray emissions.
The solar flare is a gradual hard X-ray burst (GHB) of April 26, 1981 (Cliver et al. 1986).
In many cases GHBs  are preceded by coronal mass ejection and with a hard X-ray peak.
The flux units are in relative units, while each time unit is 1 hour for the solar
flare, and 7.1 weeks for Aql~X-1. Namely, a ratio of 170 in the time scale.
Radio fluxes are in thin lines and X-ray fluxes are in thick lines.
Aql~X-1 is depicted by a blue dashed line, while the solar flare is
shown with a red solid line.
{{{{{ Note that the Aql~X-1 radio intensity was multiplied by 600, as the ratio $L_r/L_x$ is
$\sim 600$ times weaker in Aql~X-1 as in the solar flare that is shown. }}}}}

I note the following similarities between the bursts in these two vastly different systems:
\begin{enumerate}
\item The general shape of the radio (3.6~cm-6~cm) and the soft ($2-10 \keV$) X-ray emissions is similar,
with a ratio of $\simeq 170$ in time scale between the two systems.
\item The beginning of the soft X-ray emission appears before the radio.
 Part of this early rise might be a contribution from the hard X-ray component.
However, from the different behavior of the two components, some of the early rise is
part of the soft X-ray component itself.
\item However, the steep rise in X-ray flux comes only after the radio peak.
\item The rise to maximum in radio flux is slightly faster than the decline.
\item In the 1981 April 26 solar GHB there is a hard X-ray peak preceding the radio
       peak (fig. 1 in Cliver et al. 1986; fig. 7 in Dennis \& Zarro 1993).
        This is similar to the hard X-ray peak preceding the radio peak in Aql~X-1.
        However, I do not claim the same explanation for the hard X-ray peaks in the
        two systems. I only claim that the hard X-ray emission is connected to the process
        that powers the radio and soft X-ray peaks that follow.
\end{enumerate}

{{{{{   There is a weaker second radio peak in the light curve of Aql~X-1.
The popular accretion disc-radio jet paradigm (the spectral state transition paradigm)
predicts reflaring in radio as the source hardens.
The magnetic flare interpretation proposed here, on the other hand,
requires the presence of a second flare.
Indeed, the small radio peak precedes a major X-ray flare that has its X-ray peak
24 days after the second radio peak (Swift/BAT Hard X-ray Transient Monitor home-page).
It is possible that a larger radio peak was presence closer to the second X-ray peak, but
no radio data exit.
Consecutive flares are seen in stars.
The January 7 1995 Flare C of UV Ceti occurred during the decline of Flare B (G\"udel et al. 1996).
The radio peak of Flare C was stronger than that in Flare B, but its X-ray emission was much weaker. }}}}}

\subsection{Fluxes Ratio}
\label{sec:flux}
LB2008 present a correlation between the radio and soft X-ray luminosities
$L_r/L_x  \simeq 10^{-5}$, that holds over 20 orders of magnitude,
excluding some systems, e.g., the Sun, Galactic BH XRBs,
and radio loud quasars.
In some cases this ratio is as low as $\sim 10^{-8}$, in particular in weak solar flares
(micro-flares; Benz \& G\"udel 1994) and in Galactic BH XRBs.
{{{{{ LB2008 raised the possibility that the lower $L_r/L_x$ in BH XRBs
is due to the higher temperature of their disk. If this is indeed the case,
it strengthens the connection between the flaring systems proposed here. }}}}}

In the November 2009 outburst of Aql~X-1 this ratio was $(L_r)_{\rm peak}/(L_x)_{\rm peak}  \simeq 10^{-8}$.
At the radio peak the X-ray luminosity was lower, and this ratio was $\sim 3 \times 10^{-8}$.
This is similar to weak solar flares.
In the solar flare from April 26, 1981 this ratio is $L_r/L_x \sim 10^{-5}$
(the total energy in the radio is six orders of magnitude below that in the X-ray).
We note that the ratio in solar flares is mainly obtained with the soft X-ray flux
$0.5-2 \keV$ (e.g., G\"udel et al. 1996).
Taking the $0.1-20 \keV$ range for the solar flare, will reduce the $L_r/L_x$
ratio only by a factor of 2.
The fluxes for Aql X-1 is taken for the $2-16 \keV$ band (M2010), which is somewhat narrower
than the band used by LB2008 ($0.2-20 \keV$). For the spectra of these sources
this does not make a significant difference, as emission peaks around a few keV.

In accretion onto compact objects some X-ray emission comes from shocked gas, a process that does
not exist in the Sun.
Over all, the ratio $L_r/L_x  \simeq 10^{-8}$ in the November 2009 outburst of Aql~X-1
falls within the range of weak solar flares and BH XRBs too.

\subsection{Flat radio flux}
\label{sec:radioflux}

{{{{{ M2010 attribute the flat radio spectrum to an optically thick compact jet.
However, in BH XRB outbursts,
the radio emission after a major radio flare is dominated by optically
thin jet ejecta, and there is currently no evidence that the compact jet remains on. }}}}}
For the November 2009 outburst M2010 find (near 8~GHz)
$S_\nu({\rm radio}) \propto \nu^{ 0.40 \pm 0.31}$.
The weighted VLA data yields $\propto \nu^{0.05 \pm 0.16}$.
Both measurements would need to be off by 3 sigma to be consistent with $\nu^{-0.5}$
{{{{{ (as in optically thin jet ejecta). }}}}}
According to M2010, this is not compatible with internal shocks in jets.
It thus seems that a magnetic flare can account for this radio spectrum.
Kundu et al. (2009), for example, find for the magnetic looptop in a solar flare
$S_\nu({\rm radio}) \propto \nu^{0.5}$ near 8~GHz.

{{{{{
Although XRB systems might show different behaviors (e.g., Agrawal \& Misra 2009),
in general XRB emission can be well fitted by a combination of a soft component
composed of a multi-color disc blackbody and a hard component of
a power-law or Comptonized single temperature (e.g., Maitra \& Bailyn 2004;Gou et al 2009).
Stellar (including solar) flares can be fitted by emission from thin plasma (coronal emission),
e.g., Battaglia et al. (2009).
The X-ray spectra of stellar flares, therefore, are not similar to those of XRB systems.
But they do not need to be identical in the proposed model.
LB2008 note the different X-ray spectra of AGN and stellar flares.
They argue that the cooling mechanism of the hot gas formed by the magnetic activity is different.
While in stellar flares the cooling is by thermal emission from optically thin plasma, in
AGN the cooling is Comptonization of the optical-UV disk continuum by hot thermal gas
($T \ga 10^9 \K$). Still, LB2008 argue, in both types of systems the hot plasma is
formed by magnetic activity.

As for XRB systems discussed here, it is possible that the hot gas is cooling by heating the disk.
For example, the hot electrons formed in the magnetic reconnection process stream down
along magnetic field lines anchored to the disk.
This will increase the black body emission from the disk.
In addition, some contribution from optically thin plasma (as in stellar flares) is expected.
These processes are the subject of a future paper.
In particular two processes will be studied: The formation of the very hot gas by magnetic activity
in the flaring region, and heat conduction from that region to the disk.
After all, this chain of event is what is behind the Neupert effect: hot electrons heat (and evaporate)
the chromosphere.
This future study will have to examine where the heat conduction in the environment of accretion disks
can indeed heat the disk to increase its blackbody radiation.
 }}}}}

\section{DISCUSSION AND SUMMARY}
\label{sec:summary}

The three properties discussed in Section \ref{sec:flare} hint that the behavior
of the soft X-ray and radio emissions in the November 2009 outburst of the XRB
Aql~X-1, were similar to gradual hard X-ray bursts (GHBs)
in the Sun and similar stars, e.g., UV Ceti (G\"udel et al. 1996).
The strongest of these properties is the similar light curve in hard X-ray,
soft X-ray, and radio emissions.
The similarity in the radio and soft X-ray light curves is shown in Figure \ref{fig:flare}.
The hard X-ray peak in Aql X-1 can be attributed to the disk instability (Dubus et al. 2001).
It is the behavior of the soft X-ray and radio emissions that I try
to explain in this Letter.

The typical size of the soft X-ray loop of solar GHBs is
$\sim 0.04 R_\odot$ (see fig. 10 in Cliver et al. 1986).
If we use a very simple scaling from the solar flare, the duration
of the Aql~X-1, which is $\sim 170$ times as long as a solar flare, implies a reconnection
size that is $\sim 170$ times as large, or $D_{\rm rec} \sim 7 R_\odot$.
Interestingly, the orbital separation in Aql~X-1 is $4.5 R_\odot$;
for an orbital period of 19~h (Chevalier  \& Ilovaisky 1991)
and a total mass of $2 M_\odot$ (Welsh et al. 2000).
Namely, the magnetic field reconnection event occurs over a size about
equal to the binary separation, which is just a little larger than the disk size.

I propose that a disk instability triggered the formation of an extended corona
around the accretion disk, on a size about equal to the binary separation.
This is the event seen in the hard X-ray of Aql~X-1 by M2010 that preceded both
the radio and soft X-ray fluxes.
Then, a major magnetic flaring (reconnection and other means of magnetic energy release)
occurred.
Such a magnetic flare can accelerate gas and lead to the formation of a jet.

The total energy emitted in the soft X-ray in the flare is $\sim 10^{43} \erg$.
Taking the total volume occupied by the reconnecting magnetic fields to be
$V=(4 R_\odot)^3$, the required average magnetic field intensity is $B \simeq 10^5$~G.
This is not unreasonable.
Active stars with a similar size of $\sim R_\odot$ have
magnetic fields of $\sim 10^4~$~G.
The magnetic field of the Ap star HD~137509, for example, reaches
values of $37,000~$G (Mathys 1995).
{{{{{ The magnetic field comes from a disk dynamo, which should be a major issue
to be studied in the future. }}}}}

Encouraging are also some similarities of the properties mentioned in this paper
to strong and long X-ray flares of young stars characterized by having active
or inactive accretion disks around them.
Favata et al. (2005) observed the Orion Nebula Cluster with {\it Chandra},
and analyzed flares from young stars with such disks around them.
Some flares last for up to 5 days. The structure Favata et al. (2005) infer
is of large and long, $\sim 10^{12} \cm =14 R_\odot$, magnetic structures.
A large and long magnetic loop can be behind the outburst of Aql~X-1 in the proposed
model.
The total energy in the strong flares studied by Favata et al. (2005) is $\sim 10^{37} \erg$.
This is $\sim 10^{-6}$ times that in the flare of Aql~X-1.
The six (out of 32) brightest flares have luminosities in the range of
$L_x=1.6-7.8 \times 10^{32} \erg \s^{-1}$, or $\sim 10^{-5}$ times that of the
outburst of Aql~X-1.
This ratio is of the same order of magnitude as the ratio of the gravitational potential wells
in the two types of objects

Some points must be clarified before the speculative interpretation suggested here
can gain more credibility.
(1) It should be shown that a reconnection event (or several events) can account for the
   high luminosity in the X-ray (up to $\sim 0.1 L_{\rm Edd}$).
(2) There is a need to compare not only the observed properties, but also the physical processes.
  Namely, lots of work has been done on solar flares and GHBs. It will have to be shown that
  similar processes can take place in an extended coronae of XRBs.

The proposed magnetic flare explanation is a paradigm shift from the common
interpretation of outburst in XRBs.
This paradigm shift can be traced back to LB2008 who
pointed out the similar radio to X-ray fluxes ratio in a variety of astronomical objects,
including stellar flares and accreting super massive BHs in active galactic nuclei.
I therefore suggest that the proposed magnetic flaring model holds
for BH XRBs as well (see also, e.g., Rodriguez \& Prat 2008).
Basically, where we expect strong dynamo to operate, we should expect strong magnetic flares.
Accretion disks have very strong shear compared with stars, and they must have some kind
of turbulence to supply the viscosity. As such, a strong dynamo exists in accretion disks.
The large flares are expected to occur when the magnetic field energy is further amplified.
Such can be the case when the disk becomes unstable, and erupt.
The outflowing gas stretches magnetic field lines and substantially amplifies the field.
After this triggering eruption, the field lines reconnect and release the huge amount of
magnetic energy.
However, in some cases disk outbursts might not result in substantial magnetic field amplification,
and no large magnetic flare will occur.

If the similarity presented in this paper holds, it adds to previous suggestions
on similarity between coronae in active stars and accretion disks (LB2008),
and similarity in the processes of mass ejection by solar flares and the launching of jets
from accretion disks powered by magnetic fields reconnection
(de Gouveia Dal Pino \& Lazarian 2005; de Gouveia Dal Pino et al. 2010; Soker 2007).

I thank two anonymous referees for very helpful comments.
This research was supported by the Asher Fund for Space
Research at the Technion, and the Israel Science foundation.

\end{document}